\documentstyle[prl,aps,twocolumn,epsfig,floats]{revtex}

\begin{document}
\draft

%---------------------------------------------------------------
%
%  Title
%
%---------------------------------------------------------------

\title{Experimental Generation and Observation of Intrinsic Localized
Spin Wave Modes in an Antiferromagnet}

\author{U. T. Schwarz, L. Q. English, and A. J. Sievers}

\address {Laboratory of Atomic and Solid State Physics,
Cornell University, Ithaca, NY 14853-2501}

\date{May 1999}

\twocolumn[
\widetext
\begin{@twocolumnfalse}

\maketitle

\begin{abstract}   
By driving with a microwave pulse the lowest frequency
antiferromagnetic resonance of the quasi 1-D biaxial antiferromagnet
$\rm (C_2H_5NH_3)_2CuCl_4$ into an unstable region intrinsic localized
spin waves have been generated and detected in the spin wave gap.
These findings are consistent with the prediction that nonlinearity
plus lattice discreteness can lead to localized excitations with
dimensions comparable to the lattice constant.
\end{abstract}


\vspace{0.2cm}

\narrowtext

\end{@twocolumnfalse}
]

Although solitons continue to play an important role in condensed
matter physics \cite{CL95,KIK90,MS91} in the last decade it was
recognized that nonlinearity plus lattice discreteness can lead to a
different class of excitations with dimensions comparable to the
lattice constant \cite{SP95,Aub97,FW98}.  Such intrinsic localized
modes (ILMs) have been identified in a variety of classical molecular
dynamics simulations \cite{BKP90,DP93,RM98} and with macroscopic
mechanical \cite{RZE97} and electrical \cite {MBR95} models, all of
which ignore the possible role of quantum mechanics \cite{WGB96}.
Some effort has gone into identifying specific condensed matter
signatures as evidence of ILM production but all require intricate
arguments: these include far infrared absorption \cite{ST88},
radiation ionization tracks \cite{RC95}, the temperature dependent
M\"ossbauer effect \cite{SP95} and resonant Raman scattering
\cite{SBL99}.  As yet evidence of externally generated ILMs in a
lattice of atomic dimensions is missing.  The large amplitude
modulational instability of an antiferromagnetic resonance (AFMR) for
some antiferromagnets has been suggested \cite{LS98,LS98a} as a
mechanism for the generation of intrinsic localized spin wave modes
(ILSMs) and in this letter we describe the experimental observation
and control of such nanoscale excitations.

Because the driving field necessary to create ILMs via the
modulational instability scales with the frequency of the
antiferromagnetic resonance, $\omega_{\rm AFMR}$, $\rm
(C_2H_5NH_3)_2CuCl_4$ with the lowest frequency resonance at
$\omega_{\rm AFMR} = 1.5 \rm\, GHz$ was chosen for this first study.
This antiferromagnet, often referred to as $\rm C(2)CuCl_4$, is a
layered organic material with a strong ferromagnetic coupling of the
magnetic $\rm Cu^{2+}$ ions within a layer and with a weak
antiferromagnetic coupling between these layers.  Because of this weak
interlayer coupling the total spin in each layer can be represented by
a classical one with respect to describing the lowest frequency mode
dynamics.  This spin system with its biaxial anisotropy is described
in more detail in Ref. \onlinecite{LS98a}.

The $\rm C(2)CuCl_4$ single crystals were grown from aqueous solution
of ethylamine hydrochloride and copper (II) chloride in a closed
vessel by slowly decreasing the temperature \cite{JAM72}.  For the
measurement, platelets with well defined surfaces and a typical
dimension of $3\times 3 \times 0.5 \rm\, mm^3$ were chosen.

%%%%%%%%%%%%%%%%%%%%%%%%%%%%%%%%%%%%%%%%%%%%%%%%%%%%%%%%%%%%%%%%%%%%%%%
% Figure 1a, b
%%%%%%%%%%%%%%%%%%%%%%%%%%%%%%%%%%%%%%%%%%%%%%%%%%%%%%%%%%%%%%%%%%%%%%%

\begin{figure} [t!]
\centerline{\epsfig{file=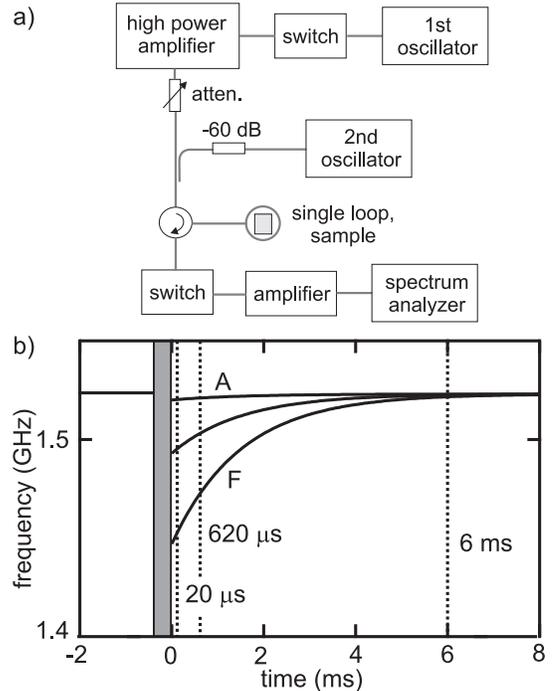,width=2.8in}}
\caption{(a) Schematic experimental setup. Details are given
in the text. (b) Experimental measurement road map. The horizontal
line for $t < 0$ identifies the AFMR frequency.  During the high power
driving pulse (shaded area) the detection electronic is blocked. The
solid lines A ... F show the AFMR for increasing pump-pulse power
levels. The vertical dotted lines identify different time-cuts
described in the text.}
\end{figure}

The experimental setup is shown schematically in Fig. 1(a). The first
oscillator provides the high power pulse and the second oscillator the
probe beam.  Because the expected experimental signature of ILSM
generation after a strong microwave pulse near the $\omega_{\rm AFMR}$
is the breakup of the AFMR into a broad band, the generation/detection
system is tailored to produce a large AC field at the driving
frequency $\omega_{\rm excite}$ and a high sensitivity over a broad
band below but close to $\omega_{\rm excite}$. A single $3\rm\, mm$
diameter loop of copper wire is used as a non-resonant antenna for the
excitation and pickup of the broad band signal in reflection. A pump
signal of up to $100 \rm\,W$ can be obtained from the oscillator and
solid state amplifier.  To create short pulses a fast GaAs switch is
used in front of the amplifier. The signal of a second, tunable
oscillator is overlayed by means of a directional coupler. With a
second fast switch, synchronized by a digital delay generator, the
reflected signal from the pump pulse is suppressed.  The weak
reflected signal from the second oscillator passes through two low
noise amplifiers and is detected with a spectrum analyzer which was
used as a variable bandwidth detector locked to the frequency of the
probe oscillator.  The minimum noise level of the system is $-120
\rm\,dBm$ at $100 \rm\,kHz$ bandwidth. In this measurement setup the
absorption can be obtained as a function of time and frequency by
first measuring the time dependent absorption at the frequency of the
second oscillator, and then scanning this oscillator frequency during
subsequent pump pulses.  The time resolution is limited to $20
\rm\,\mu s$ by the spectrum analyzer, and the sensitivity, by the
flatness of the exciting loop impedance. Care was taken not to
saturate any electronic circuits, as this could produce an artificial
nonlinear response.  The exciting coil and sample are immersed in
pumped liquid helium at $1.2 \rm\,K$.

To provide a road map to the experimental data-taking procedure a
schematic view of the power and time dependence of the AFMR is shown
in Fig. 1 (b).  The AFMR frequency before the pulse, $t < 0$, is
represented by the horizontal line.  During the high power driving
pulse (shaded area) the detection electronics is blocked.  After the
trailing edge ($t = 0$) of the driving pulse the spin wave is highly
excited and its frequency is decreased due to the intrinsic
nonlinearity of the spin Hamiltonian.  From this state the system
relaxes back to equilibrium with the longitudinal relaxation time
T$_1$. The absorption by the spin system can be examined for
different power levels of the driving pulse [lines labeled A ... F in
Fig. 1(b)] or for different time delays at a fixed power [dotted lines
in Fig. 1(b)] or as a function of the driving pulse length.
Measurements varying all of these experimental parameters have been
carried out and are described below. 

Figure 2 shows the absorption spectra $20 \rm\,\mu s$ ($\gg \rm T_2$
of the AFMR) after the trailing end of a pump pulse for six different
powers.  The power of the $400 \rm\,\mu s$ long driving pulse varied
from $50 \rm\,mW$ to $1.6 \rm\,W$. On the low frequency side of each
spectrum are two magnetostatic volume modes while the shoulder on the
high frequency side is a surface mode \cite{RHW75}. With increasing
power levels the AFMR first collapses into a broad asymmetric shape
and then at still higher powers returns to a sharp AFMR.  Note that
the magnetostatic modes do not show the same behavior as the AFMR.
These experimental results demonstrate that the AFMR is indeed
unstable with increasing amplitude but only up to a specific amplitude
while for still larger amplitudes the AFMR uniform mode again becomes
stable.  

The results in Fig. 2 illustrate some of the criteria for the creation
of ILSMs.  First there is a minimum transverse amplitude above which
the extended mode becomes unstable but then, surprisingly, there is
also a maximum amplitude associated with this instability of the
extended mode.  This was not expected from our molecular dynamics
simulations. Next the frequency interval of the instability region is
observed to become larger as the frequency shift $\omega_{\rm
excite}-\omega_{\rm AFMR}$ becomes larger.  For optimum conditions,
localization could be observed for powers as low as $50 \rm\,mW$ in a
$400 \rm\,\mu s$ driving pulse.  Varying the pulse width from $50
\rm\,\mu s$ to $400 \rm\,\mu s$ while keeping the energy in the pulse
fixed does not change these criteria.

%%%%%%%%%%%%%%%%%%%%%%%%%%%%%%%%%%%%%%%%%%%%%%%%%%%%%%%%%%%%%%%%%%%%%%%
% Figure 2
%%%%%%%%%%%%%%%%%%%%%%%%%%%%%%%%%%%%%%%%%%%%%%%%%%%%%%%%%%%%%%%%%%%%%%%

\begin{figure}
\centerline{\epsfig{file=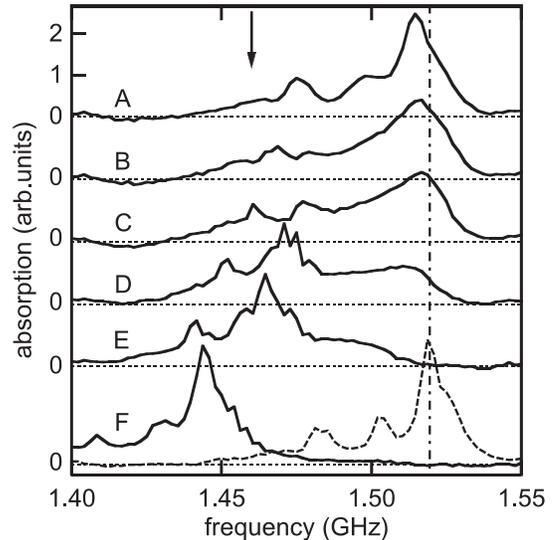,width=2.8in}}
\caption{Absorption spectra measured $20 \rm\,\mu s$ after a $400
\rm\,\mu s$ driving pulse for the different power levels. (A) $50
\rm\,mW$, (B) $100 \rm\,mW$, (C) $160 \rm\,mW$, (D) $320 \rm\,mW$, (E)
$500 \rm\,mW$, and (F) $1.6 \rm\,W$ of the $1.46 \rm\,GHz$ (arrow)
driving pulse.  For clarity the spectra are shifted on the linear
ordinate.  The asymmetric wing on the low frequency side of the AFMR
is the expected signature of localized mode production.  The
dot-dashed line marks the lowest frequency $\omega_{\rm AFMR}$ for an
applied static field $\rm H_{DC} = 12 \,mT$.}
\end{figure}

The narrow linewidth of curve F shown in Fig. 2 after the strongest
driving pulse is a handy test to exclude any temperature related
effect, such as the heating of the crystal by the intense microwave
pulses.  From low power linear measurements both the temperature
dependence of the AFMR frequency, which goes to zero at the N\'eel
temperature $\rm T_N = 10.2 \,K$, and its linewidth are known.  If the
frequency shift of $75 \rm\,MHz$ between the low power trace (dashed
line in Fig. 2) and high power trace were caused by an increased
temperature in the crystal, the line width should nearly double.  This
is not observed so temperature effects are not the source of the
unusual lineshape results.  

The frequency response at several delay times for $200 \rm\,mW$
excitation in a $200 \rm\,\mu s$ pulse is shown in Fig. 3.  The five
traces at $20 \rm\,\mu s$, $220 \rm\,\mu s$, $420 \rm\,\mu s$, $620
\rm\,\mu s$ and $6 \rm\,ms$ show the evolution of this absorption
feature from a broad resonance back to an AFMR at times smaller than
$\rm T_1 = 1.50 \,ms$.  For times larger than $\rm 2 T_1$ only the
AFMR survives.  The radical difference between the line shapes at $t =
20 \rm\,\mu s$ and $t = 420 \rm\,\mu s$ is consistent with the idea
that the breakup into ILSMs is only possible when the effects of
nonlinearity and dispersion are much stronger than the dissipation
effect \cite{LS98}.
 
%%%%%%%%%%%%%%%%%%%%%%%%%%%%%%%%%%%%%%%%%%%%%%%%%%%%%%%%%%%%%%%%%%%%%%%
% Figure 3
%%%%%%%%%%%%%%%%%%%%%%%%%%%%%%%%%%%%%%%%%%%%%%%%%%%%%%%%%%%%%%%%%%%%%%%

\begin{figure}
\centerline{\epsfig{file=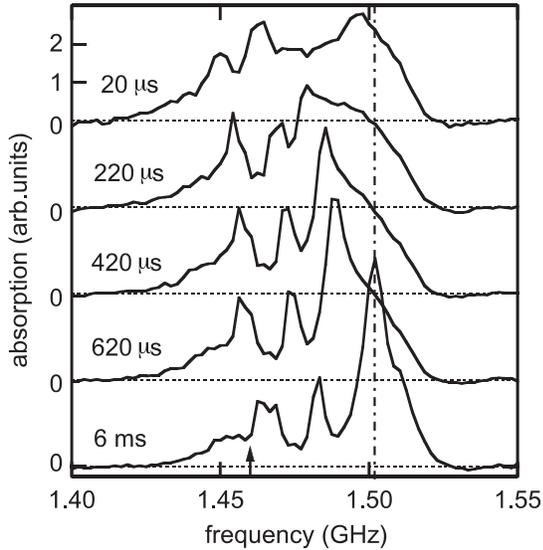,width=2.8in}}
\caption{Absorption spectra measured $20 \rm\,\mu s$, $220
\rm\,\mu s$, $420 \rm\,\mu s$, $620 \rm\,\mu s$ and $6 \rm\,ms$ after
a $200 \rm\,\mu s$ long driving pulse at $1.46 \rm\,GHz$ (arrow). The
spectra are shifted on the ordinate by equal amounts.  The dot-dashed
line marks $\omega_{\rm AFMR}$ as in Fig. 2.}
\end{figure}

MD simulations can be used to demonstrate that the extended spin wave
is unstable against breakup into localized modes for the specific
amplitude of the extended wave produced in our experiments.  Details
on molecular dynamics procedures for simulating localized spin waves
in $\rm C(2)CuCl_4$ can be found in Ref. \onlinecite{LS98a}.  The
solid line in Fig. 4(a) shows the calculated frequency dependence of
the AFMR as a function of its transverse amplitude $S_y$ in the hard
axis direction. To compare this simulation to our experiment, we
extract from the shift of $75 \rm\,MHz$ or $0.05 \,\omega_{\rm AFMR}$
of trace (F) in Fig. 2 the spin wave amplitude $S_y = 0.06$ as marked
with (F) in Fig. 4(a). Correspondingly the points A to F superimposed
on this curve identify different AFMR frequency shifts observed in the
high power measurements of Fig. 2.  To test for the instability
threshold the time dependent evolution of the energy density for a
simulation of a $250$ spin antiferromagnetic chain is calculated,
starting from an extended wave with fixed transverse amplitude plus
random noise $\langle\delta S_n\rangle=0.0025$.  For the whole range
(A) to (F) covered by the experiment the extended spin wave is
unstable and breaks up into localized modes with widths extending over
about $10$ lattice constants. Thus for these microwave powers the AFMR
is driven into the interesting nonlinear region.

%%%%%%%%%%%%%%%%%%%%%%%%%%%%%%%%%%%%%%%%%%%%%%%%%%%%%%%%%%%%%%%%%%%%%%%
% Figure 4a, b
%%%%%%%%%%%%%%%%%%%%%%%%%%%%%%%%%%%%%%%%%%%%%%%%%%%%%%%%%%%%%%%%%%%%%%%

\begin{figure}
\centerline{\epsfig{file=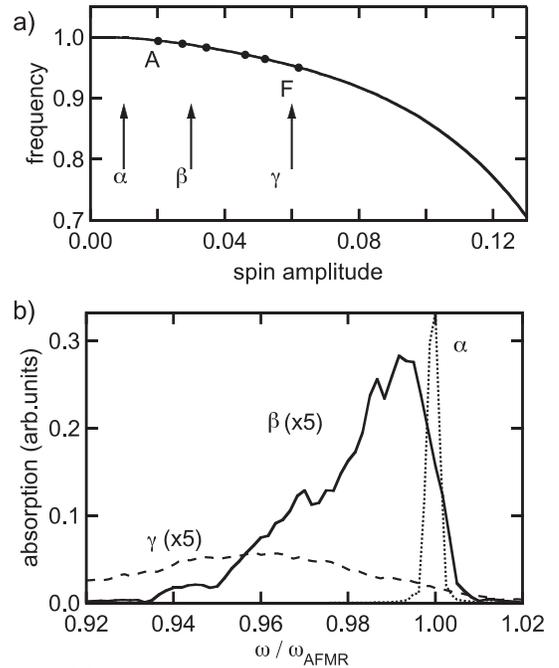,width=2.8in}}
\caption{(a) Calculated frequency of the uniform spin wave mode as
function of its transverse amplitude. (A) to (F) mark the
experimentally determined position of the extended mode after the
driving pulse from the corresponding traces in Fig. 2. (b) AFMR plus
ILSM absorption spectra determined from molecular dynamics simulations
corresponding to three different microwave power levels.  Starting
from extended spin waves with a given transverse amplitude of $Sy =
0.01$, $0.03$ and $0.06$ (arrow positions $\alpha$, $\beta$, $\gamma$
in Fig. 4a) the uniform mode breaks up into ILSMs within a time of
$200 \,\tau_{\rm AFMR}$.  From the evolution over the subsequent time
interval $\Delta\tau= 800 \,\tau_{\rm AFMR}$ the spectra are obtained
via the auto-correlation of the net transverse magnetic moment as
described in the text.}
\end{figure}

To compare the absorption spectrum $A(\omega) \propto \omega
\chi^{\prime\prime}$ measured in the experiment with results from
molecular dynamic simulations, the imaginary part of the dynamic
magnetic susceptibility is calculated using the Kubo
expression\cite{KT54} where the absorption is proportional to the
Fourier transform of the auto-correlation function of the net magnetic
moment, $M_y(t)$:

\begin{equation}  
\chi_y^{\prime\prime}(\omega)\propto\omega\int_0^\infty dt \langle
M_y(t_0+t)M_y(t)\rangle \exp(i\omega t).
\end{equation}   

For our special case the simulations are started with an extended wave
of a given amplitude plus some small amount of random noise.  The
extended wave is unstable against breakup into localized excitations
and transforms after approximately $100$ periods of the
antiferromagnetic resonance into a broad spectrum of ILMs.  From $200$
to $1000 \tau_{\rm AFMR}$ the evolution of the net magnetic moment is
recorded and then via Eq. 1 the imaginary part of the dynamic magnetic
susceptibility is found.

The resulting calculated absorption spectra for a chain of $1000$
spins at three different powers are shown in Fig. 4(b).  To generate
these spectra the starting transverse amplitudes of the AFMR are $Sy =
0.01$, $0.03$, and $0.06$, respectively.  Each curve is averaged over
6 simulation runs to remove arbitrary spikes associated with single
long-living ILSMs.  The broad wing of the asymmetric band is due to
the statistical distribution of the ILMs with different degrees of
localization and thus different frequencies.  To compare the
simulations on the long chain with the spectra obtained in the
experiment, the frequency axis in Fig. 4(b) was scaled to the same
width relative to $\omega_{\rm AFMR}$ as given in Figs. 2 and 3.  Both
the shape and width of the experimental onset results are in good
agreement with theoretical simulations.  Moreover the absorption
spectra shown in Fig. 4(b) are obtained starting from amplitudes of
the extended wave which are in the right range for traces A to F.

The disappearance of the instability at large amplitudes is an
experimental feature not represented by curve $\gamma$ in these model
calculations.  Since the values of the relaxation times taken from the
experiment are $\rm T_1 = 1.50 \, ms$ and $\rm T_2 \approx 0.1
\rm\,\mu s$, with the latter depending on the surface quality of the
samples, T$_2$ is much smaller than the pulse widths used in our
experiments so that energy is transferred to other degenerate spin
waves during the microwave pulse.  Our experimental results indicate
that at the powers where the uniform mode instability occurs this
transfer does not influence the positive curvature of the dispersion
curve which is required for the instabilty.  At the highest powers
shown in Fig. 2 the uniform mode again becomes stable indicating that
the finite wavevector spin waves have become so heavily populated
during the microwave pulse that the dispersion curve now has negative
curvature at the uniform mode, a condition for stability of the mode
at large amplitude \cite{LS98}.  Because it takes a finite time for
the uniform mode to break up into ILSMs both features can appear at
intermediate powers such as displayed in trace D in Fig. 2.  Here
ILSMs are formed during the early part of the pulse while the
population in the degenerate modes is still small, and they remain
isolated in frequency space above the dispersion curve after the
degenerate mode population becomes large so both features are seen.
	
In this series of microwave experiments the instability that appears,
when the lowest lying AFMR of $\rm C(2)CuCl_4$ is driven to larger
amplitudes, has been used to generate nonlinear excitations which are
localized on a nano-length scale.  To distinguish between the ILSMs of
classical simulations and the excitations observed in experiment the
latter will be identified as \lq anons\rq.  The hallmark experimental
feature of the uniform mode breakup into anons is a broad and
asymmetric spectral band below the AFMR.  Classical MD simulations on
a long 1-D antiferromagnetic spin chain show that both the amplitude
of the extended spin wave at which it becomes unstable to breakup into
localized modes and the resulting spectral shape of the ILSMs are in
good agreement with the microwave results.  At still higher
experimental powers the AFMR is observed to become stable again, a
feature related to the fact that the pump pulse is longer than T$_2$.

Simulations have played an important role in identifying the most
economical experimental pathway for the detection of anons in real
solids.  This interplay may be expected to continue since additional
MD studies on a $\rm C(2)CuCl_4$ spin chain in the presence of a
magnetic field gradient indicate directed anon transport should be
possible.

The authors thank H. Padamsee who provided the microwave amplifier for
these experiments and also R. Lai and R. H. Silsbee for helpful
discussions.  This work is supported by NSF-DMR-9631298.  One of the
authors (U. T. S.) is supported in part as a Feodor-Lynen scholar by
the Alexander von Humboldt-Foundation.

%---------------------------------------------------------------
%
%  References
%
%---------------------------------------------------------------

%\begin{references}

%\bibliographystyle{prlsty}
%\bibliography{paper}    

\end{document}